\documentstyle[12pt,epsf]{article}
\begin{document}

\title{Effects of nuclear structure on average angular \\
 momentum in subbarrier fusion}

\author{{\underline {A.B. Balantekin}}, J.R. Bennett \\
Physics Department, University of Wisconsin, \\
Madison, Wisconsin, 53706 USA\\
\and
S. Kuyucak \\
Department of Theoretical Physics,\\ Research School of Physical Sciences,\\
Australian National University,
Canberra, ACT 0200, Australia\\}
\date{}

\maketitle

\begin{abstract}
We investigate the effects of nuclear quadrupole and hexadecapole
couplings on the average angular momentum in sub-barrier fusion reactions.
This quantity
could provide a probe for nuclear shapes, distinguishing between
prolate vs. oblate
quadrupole and positive vs. negative hexadecapole couplings.
We describe the data in the O + Sm
system and discuss heavier systems where shape effects
become more pronounced.
\end{abstract}

\noindent
Address correspondence to A. B. Balantekin.

\noindent
E-mail: baha@wisnud.physics.wisc.edu
\vfill \eject

Enhancement of subbarrier fusion cross sections in medium and heavy mass
nuclei due to coupling of internal degrees of freedom to the relative motion
is a well established phenomenon (see Refs. \cite{beck} and \cite{vand} for
reviews).
Because subbarrier fusion is sensitive to details of nuclear structure, one
can consider using it as a technique to measure structure information which
is complementary to other methods. As is well known, the so-called
``distribution of barriers'' \cite{rowley}, determined from the quantity
\begin{equation}
D (E) \equiv \frac{d^2}{dE^2} (E \sigma (E)),
\end{equation}
has a shape which distinguishes among different types of nuclear structure,
for example prolate versus oblate, rotational versus vibrational, etc. In this
paper, we point out that the average angular momentum transferred to the
compound nucleus, $\langle\ell\rangle$, is also sensitive to different kinds of
nuclear shapes.
Also, we emphasize that partial cross section distributions $\sigma_{\ell}$
having
different mean values $\langle\ell\rangle$ can give the same total cross
section $\sigma$. Therefore, it is important to consistently describe both
the fusion cross section and the average angular momentum data.

For several systems, a ``bump'' in
$\langle\ell\rangle$,
a sudden rise in the angular momentum as the energy increases
toward the
barrier from below, followed by a flattening off above the barrier,
has been observed
\cite{silicon}.
We will show that
the ``bump'' in the $\langle\ell\rangle$-distribution happens for prolate
systems but not for oblate ones.
Other authors have previously noted this difference in the
$\ell$-distributions of a prolate and an oblate nucleus \cite{row2}. We
emphasize here that the different distributions are characteristic of the two
shapes.
Also,  we show that the addition of a positive or negative
hexadecapole moment can add or subtract, respectively, several units of
angular momentum from the bump. Hence, the bump in $\langle\ell\rangle$, or its
absence, can be a signature for various nuclear deformations.
In systems with
light projectiles such as $^{16}$O, observing such
signatures requires measuring $\langle\ell\rangle$ to within one unit of
angular momentum which may not be easy. Using heavier projectiles, e.g.
$^{40}$Ca, magnifies the coupling effects several times, which offers
an easier identification of nuclear shape effects.

We calculate the total fusion cross section using the partial wave expansion
in the barrier penetration picture.
The penetration probabilities for the different partial
waves are evaluated numerically using a uniform WKB approximation,
valid for energies both above and below the barrier
\cite{brinkbook}

\begin{equation}
T_{\ell}(E)=\sum_i \omega_i
\left[1 + exp\left(2\sqrt{2\mu\over \hbar^2}\int_{r_1}^{r_2}dr
(V(r, \lambda_i)-E)^{1\over2}\right)\right]^{-1}.
\end{equation}

\noindent
Here, $r_1$ and $r_2$ are the classical turning points of the motion, and
$\omega_i$ refers to the weight in the eigenchannel $i$ with corresponding
eigenvalue $\lambda_i$.
This result
is the familiar representation of the subbarrier fusion cross section
as a weighted average of cross sections due to tunnelling through a
set of potential barriers. In the geometrical
description of the nucleus, the barriers are determined by averaging over fixed
orientations of a deformed nucleus
\cite{chase} or over zero-point vibrations of the surface
of a vibrational nucleus\cite{zpm}.
Alternatively, in the interacting boson model
description of nuclei\cite{ibm}, the barriers are determined
as described in Ref. \cite{fus5} from
the eigenvalues
$\lambda_i$
of the coupling operator
\begin{equation}
\hat O = v_2 {Q \over \langle 2_1 ||Q||0_1  \rangle}\cdot Y^{(2)}(\hat{\bf r})
 + v_4 {H \over \langle 4_1 ||H||0_1  \rangle}\cdot Y^{(4)}(\hat{\bf r}),
\end{equation}
\noindent
where
$Q=[s^{\dagger}{\tilde d}+d^{\dagger}s]^{(2)} + \chi
[d^{\dagger}{\tilde d}]^{(2)}$
and
$H=[d^{\dagger}{\tilde d}]^{(4)}$
are the quadrupole and hexadecapole transition operators, respectively.
Here, the operators $s^{\dagger}$($s$) and $d^{\dagger}$($d$) create
(annihilate) bosons of angular momentum zero and two, respectively,
representing
correlated pairs of valence nucleons.
The parameters $v_2$ and $v_4$ describe the
quadrupole and hexadecapole couplings and are roughly proportional to
the E2 and E4 excitation strengths in the target nucleus.
In the IBM description, the weights $\omega_i$ are given by the overlaps of the
eigenstates of the coupling operator $\hat O$ with the ground state of the
target nucleus,
which we project from an
intrinsic state as follows

\begin{equation}
| J = 0, M = 0 \rangle = \frac{1}{\cal N}\int d\theta sin\theta
R(\theta) (b^{\dagger})^N |0\rangle,
\end{equation}

\noindent
where $\cal N$ is a normalization factor and
we also introduced the intrinsic boson operator,
$b^{\dagger} = x_0 s^{\dagger} + x_2 d^{\dagger}_0$.
The target wave functions are represented by the mean fields
$x_0$ and $x_2$ for the $s$ and
$d$ bosons which are determined from a given IBM Hamiltonian
(which fits the low-lying spectroscopic data) by variational techniques
\cite{1N}. The parameter $\chi$
is also determined from this method.
We emphasize here the utility of the IBM description of collective nuclear
structure in providing a description of rotational, vibrational and
transitional nuclei within a unified theoretical framework. This quality
becomes especially useful in describing the transitional
nuclei, such as $^{150}$Sm (considered later in this paper), which are
more difficult to handle in geometrical models.
As has been emphasized by many authors
\cite{nag,row}, the
validity of expression (5) depends only on the usual adiabatic and rotating
frame approximations which we have discussed previously \cite{fus3}.

For demonstrative purposes, we first consider a fictitious target nucleus
with $A=170$ and $Z=70$, to which we assign various static deformations in
order to isolate specific nuclear structure effects. In addition, we suppose
for simplicity that the projectile is $^{16}$O so that the parameters of the
Woods-Saxon potential are fixed from our previous systematic study of
subbarrier
fusion cross sections and barrier distributions with that
projectile\cite{fus5}.
Hence we take a simple global
parameterization of the Woods-Saxon potential, taking in all cases
$R_2  = 2.62 $~fm, $R_1 = 1.04 A_1^{1/3}$~fm,
$V_0 = 67.5 (A_1/144) ^{1/3}$~MeV, $a = 1.22 $~fm.

In Fig. [1a], we show for a prolate-deformed nucleus the effects of positive
and negative hexadecapole moments on $\langle\ell\rangle$. The solid curve
has no hexadecapole
moment ($v_4$ = 0), the dashed one has a positive hexadecapole moment and the
dotted curve has a negative one. The barrier for this system is at about 66
MeV.
At energies near the barrier, the positive
(negative) hexadecapole moments add (subtract) about one unit of angular
momentum. This effect is due to constructive (destructive) interference between
the quadrupole and hexadecapole couplings and was observed previously in
experiments on $^{16}$O + $^{154}$Sm, $^{186}$W subbarrier fusion \cite{ref}
and in our
systematic study of subbarrier fusion cross sections \cite{fus5}.
The effects
of hexadecapole moments on angular momentum transfer in subbarrier fusion
reactions have not been considered previously.
Fig. [1b] compares the angular momentum transfer as a
function of energy for a prolate target with that for an oblate target of the
same deformation. For reference, we also show the calculated average angular
momentum for a nucleus with no deformation.
In comparing Figs. 1a and 1b, we see that including hexadecapole coupling
changes $\langle\ell\rangle$ in the vicinity of the barrier energy by about
25 \% of the increase due to including quadrupole coupling only.
The prolate target displays the ``bump'' in the $\ell$-
distribution.
The oblate nucleus shows an
increased angular
momentum from the spherical nucleus at energies near the barrier, but does not
have the bump.

In Figs. [2a-2d], we show the results of using our global nuclear potential to
calculate the fusion cross section and average angular momentum for a series of
four isotopes, $^{148,150,152,154}$Sm, for which experimental data on the
fusion cross section are available
and average angular momenta are either measured or deduced
\cite{stokstad,jack1,jack2,jack3,wuosmaa,bierman,baba}.
The quadrupole and hexadecapole coupling strengths, $v_2$ and $v_4$, are
determined from a simultaneous fit to all the available data.
These parameters and the mean fields $x_0$ and $x_2$ for all of the
calculations
are listed in Table I.
In our previous analysis of fusion cross sections and
barrier distributions \cite{fus5} for the nuclei $^{148}$Sm and $^{154}$Sm, we
obtained values for these coupling
parameters which were mostly consistent with the known E2 and E4 matrix
elements for
those nuclei. In that study, we did not attempt to describe the data for
$^{150}$Sm and $^{152}$Sm since barrier distributions were not available for
those systems. In the present calculations, we have found that in order to
explain the fusion cross section data it is necessary to assume quadrupole and
hexadecapole couplings which do not change from nucleus to nucleus in a manner
consistent with E2 and E4 matrix elements, or equivalently with the geometrical
deformation parameters $\beta_2$ and $\beta_4$ as measured in Coulomb
excitation
experiments \cite{nds}. This behavior of the couplings
in subbarrier fusion has been observed by other authors
\cite{argentina,stokgross}.
In particular, in Coulomb excitation measurements, the nuclear deformations are
found to decrease rapidly and suddenly in the region of the phase transition
from rotational to vibrational nuclei around $^{150}$Sm
($\beta_2(^{152}\rm{Sm})=0.29, \beta_2(^{148}\rm{Sm})=0.14$),
whereas to describe
the subbarrier fusion data these deformations must change much more slowly.
This discrepancy between the deformations determined in the two different
experiments suggests that the meaning of the couplings in subbarrier fusion is
not necessarily the same as that suggested by a simple geometrical picture.
Investigation
of this point may require a description of subbarrier fusion in a deeper
theoretical framework than the one we are using and so will be left to later
studies.

In general one sees that the fusion cross sections can be described quite well
by our global model. The angular momentum distributions are also reproduced
well
except for a systematic underestimation of the angular momentum at energies
where the ``bump'' in the calculated distributions flattens off. The reason for
this discrepancy may be due to the need for inclusion of couplings to other
reaction channels or to some more basic shortcoming of our model of the fusion
process.

Our purpose in this letter is to show how nuclear structure effects can affect
the average angular momentum transferred to the compound nucleus in subbarrier
fusion reactions. We have demonstrated the sorts of effects one can expect to
observe and shown the validity of our model for reactions with heavy collective
nuclei. In order to emphasize the usefulness of subbarrier fusion
as a sensitive probe of nuclear structure, we would next like to calculate the
expected angular momentum distributions for reactions with $^{192}$Os and
$^{194}$Pt. The former nucleus is known to be prolate, the latter to be oblate.
These nuclei have never been studied in subbarrier fusion reactions, although
experiments to measure the cross section and barrier distribution are currently
being carried out at the Australian National University and the University of
Washington. In Fig. [3a] we show the average angular
momentum calculated using our global nuclear potential for the reactions
$^{16}$O + $^{192}$Os, $^{194}$Pt. The presence or
absence of the bump near the barrier indicates the difference between
prolate and oblate target. This effect can be enhanced, however, by using a
heavier projectile nucleus so that the reduced mass of the system is larger.
This dependence of the size of the bump on
the reduced mass of the entrance channel has been demonstrated in measurements
of the reactions $^{16}$O + $^{166}$Er and $^{28}$Si + $^{154}$Sm which lead to
the same compound nucleus \cite{vand}. Hence in Fig. [3b] we show
calculations for the angular momentum for the reactions
$^{40}$Ca + $^{192}$Os, $^{194}$Pt. In Fig. [3b] one can clearly see the bump
for the prolate target and no bump for the oblate one. For these calculations
with $^{40}$Ca as the projectile, we used the same global Woods-Saxon potential
as with the $^{16}$O projectile, except that the nuclear potential strength
was parameterized as $V_0 = 123.9 (A_1/144) ^{1/3}$~MeV.
Our
systematic study of reactions involving $^{16}$O projectiles gives us some
confidence in predicting the results of other reactions with that nucleus.
It is possible, however, that one must consider additional channels such as
projectile excitation in reactions with heavier projectiles.
Hence, we should emphasize that
these calculations are intended only to give a qualitative estimate of the
cross section and angular momentum to be expected for these reactions.

We have given a consistent description of subbarrier fusion cross section and
angular momentum data for a series of reactions of $^{16}$O with samarium
isotopes. The nuclear structure of the target nuclei greatly influences both
the cross sections and the angular momenta. We have demonstrated that
measurement of the
average angular momentum as a function of energy may be able to distinguish
between
target nuclei which are prolate (these show a bump near the barrier) and
oblate (these show no bump). Nuclear hexadecapole moments also influence the
angular momentum, though not as much as do quadrupole moments. We hope that our
work will prompt further experimental investigation of angular momentum in
subbarrier fusion, as this quantity provides an important constraint on
models of fusion.

\bigskip

This research was supported in part by the U.S. National Science Foundation
Grants No. PHY-9314131 and INT-9315876, in part by the Australian Research
Council and in part
by an exchange grant from the Department of Industry, Technology and Commerce
of Australia. One of us (J.R.B.) would like to thank the members of the
Department of Theoretical Physics at the Australian National University for
their hospitality while this work was being carried out.

\bigskip

\vfill \eject

\bigskip
\bigskip
\centerline{\bf Table I}
\bigskip
\begin{tabular}{lccccc}
\hline
{\em Nucleus}
& {\em $x_0$} & {\em $x_2$} & {\em $\chi$} & {\em $v_2$} & {\em $v_4$} \\
\hline
$^{148}$Sm & 0.92 & 0.39 & -1.15 & 0.18 & 0.0 \\ \hline
$^{150}$Sm & 0.91 & 0.41 & -1.15 & 0.22 & 0.02 \\ \hline
$^{152}$Sm & 0.77 & 0.64 & -1.15 & 0.25 & 0.055 \\ \hline
$^{154}$Sm & 0.74 & 0.67 & -1.15 & 0.26 & 0.06 \\ \hline
$^{170}$Yb & 0.64 & 0.77 & -0.66 & - & - \\ \hline
$^{192}$Os & 0.69 & 0.72 & -0.20 & 0.19 & -0.03 \\ \hline
$^{194}$Pt & 0.69 & -0.72 & 0.15 & 0.17 & -0.03 \\ \hline
\end{tabular}
\\
\bigskip
\bigskip
\noindent
Nuclear structure parameters for the nuclei which we consider.
\vfill \eject

\bigskip
\centerline{\bf Figure Captions}
\bigskip
\noindent
FIG. 1. a) The effect of hexadecapole coupling on average angular momentum
for a prolate target ($v_2 = 0.24$).
The dotted, dashed and solid lines are for $v_4 = -0.033, 0, +0.033$,
respectively. b) Comparison of average angular momentum for prolate (solid
line) and oblate (dashed line) targets and for no coupling (dotted line)
There is no hexadecapole coupling ($v_4 = 0$).

\bigskip
\noindent
FIG. 2. a) Comparison of predicted fusion cross section and average angular
momentum with the experimental data for the reaction $^{16}$O + $^{148}$Sm.
The coupling parameters are given in Table I. b)
 Same as Fig. 2a but for $^{150}$Sm. c) Same as Fig. 2a but for $^{152}$Sm.
d) Same as Fig. 2a but for $^{154}$Sm.

\bigskip
\noindent
FIG. 3. a) Predicted fusion cross sections and average angular momenta for
the reactions $^{16}$O + $^{192}$Os (solid line) and $^{16}$O + $^{194}$Pt
(dashed line). b) Predicted fusion cross sections and average angular momenta
for the reactions $^{40}$Ca + $^{192}$Os (solid line) and $^{40}$Ca +
$^{194}$Pt
(dashed line).

\epsfysize=7in \epsfbox{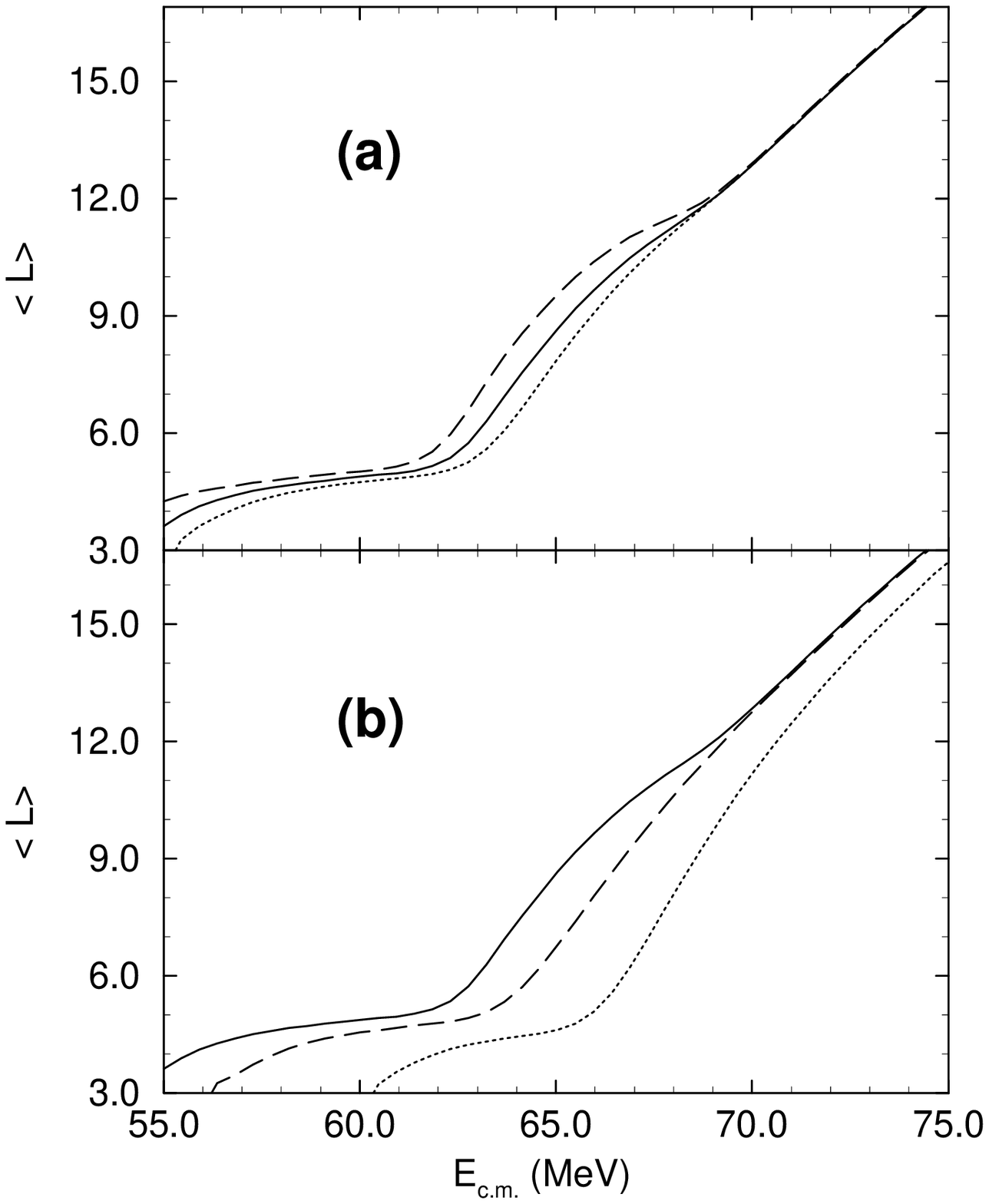}
\epsfysize=7in \epsfbox{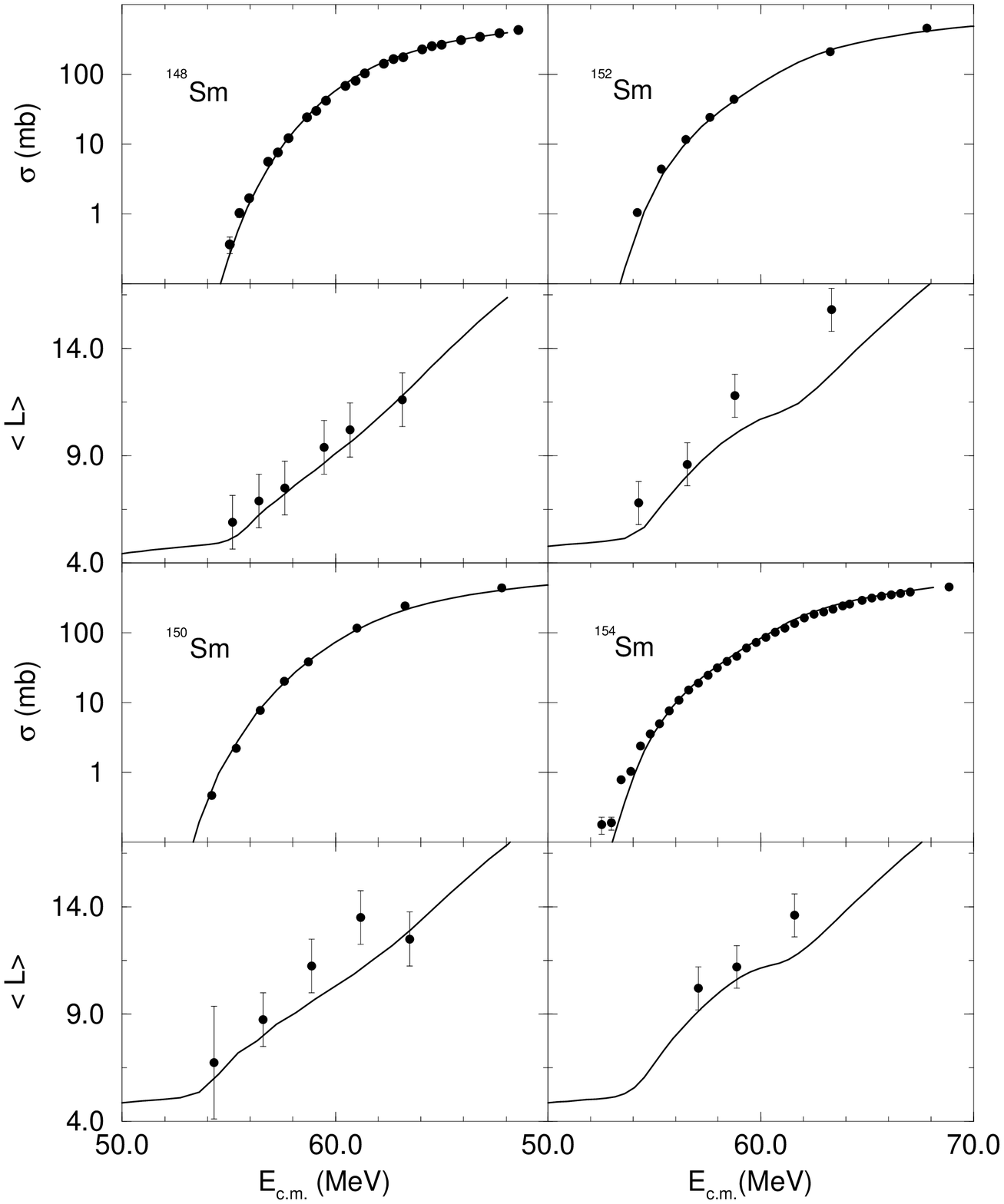}
\epsfysize=7in \epsfbox{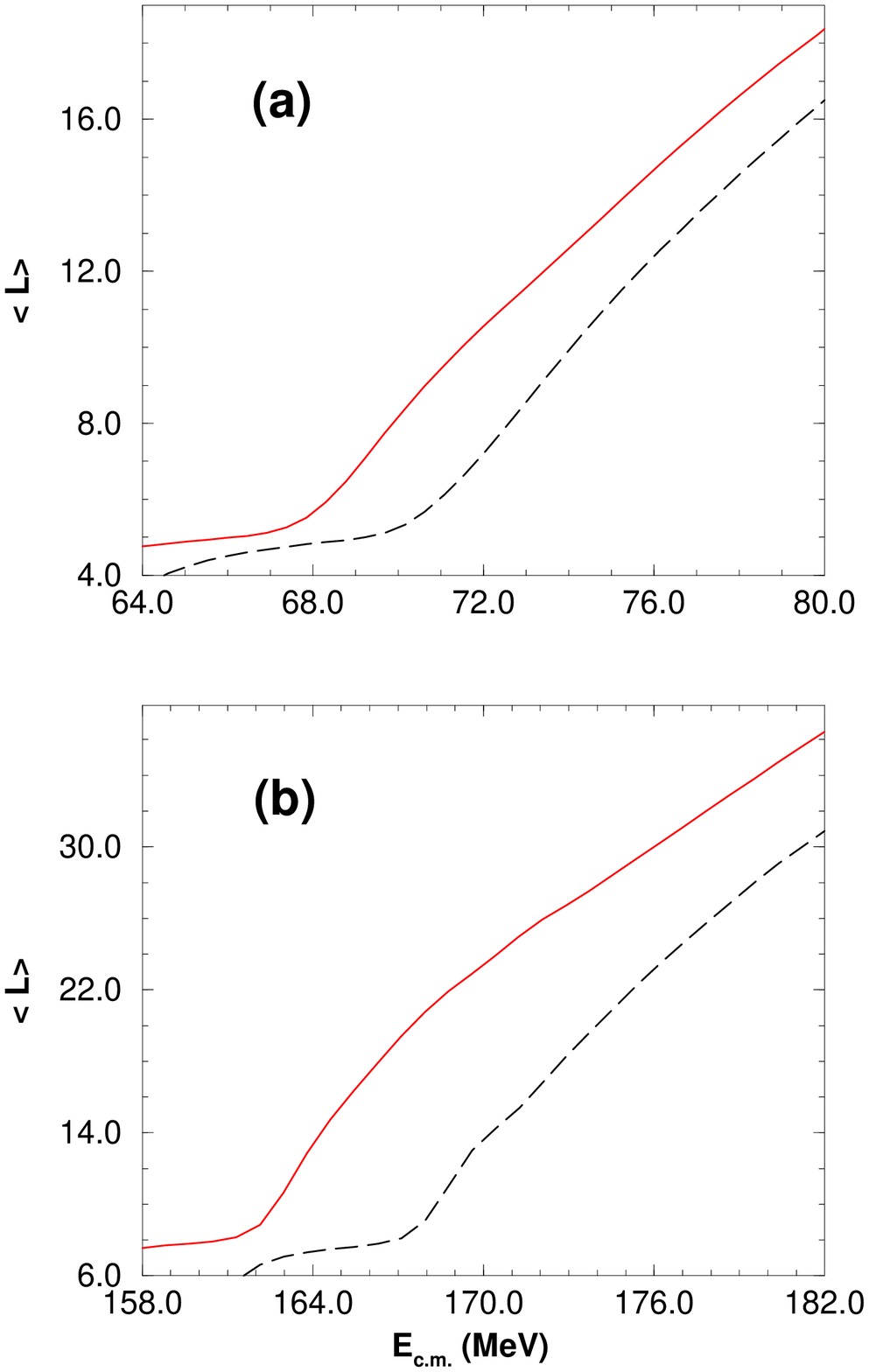}


\noindent

\begin{thebibliography}{99}

\bibitem{beck}
M. Beckerman, Rep. Prog. Phys. 51 (1988) 1047.

\bibitem{vand}
R. Vandenbosch, Ann. Rev. Nuc. Part. Sci. 42 (1992) 447.

\bibitem{rowley}
N. Rowley, G.R. Satchler and P.H. Stelson, Phys. Lett. B 254 (1991) 25.

\bibitem{silicon}
S. Gil {\it et al.}, Phys. Rev. Lett. 65 (1990) 3100.

\bibitem{row2}
N. Rowley, J.R. Leigh, J.X. Wei and R. Lindsay, Phys. Lett. B 314
(1993) 179.

\bibitem{brinkbook}
D.M. Brink, Semi-classical Methods for Nucleus-Nucleus Scattering
(Cambridge, 1985).

\bibitem{chase}
D. M. Chase, L. Wilets and A. R. Edmonds, Phys. Rev. 110 (1958) 1080.

\bibitem{zpm}
H. Esbensen, Nucl. Phys. A 352 (1981) 147.

\bibitem{ibm}
F. Iachello and A. Arima, The Interacting Boson Model (Cambridge,
1987).

\bibitem{fus5}
A.B. Balantekin, J.R. Bennett, and S. Kuyucak, Phys. Rev. C 49 (1994) 1079.

\bibitem{1N}
S. Kuyucak and I. Morrison, Ann. Phys. (N.Y.) 181 (1988) 79.

\bibitem{nag}
M. A. Nagarajan, A. B. Balantekin and N. Takigawa, Phys. Rev.
C 34 (1986) 894.

\bibitem{row}
N. Rowley, Japanese J. Appl. Phys. Series 9 (1993) 218.

\bibitem{ref}
R. C. Lemmon {\it et al.}, Phys. Lett. B 316 (1993) 32.

\bibitem{fus3}
A.B. Balantekin, J.R. Bennett, A.J. DeWeerd and S. Kuyucak, Phys. Rev.
C 46 (1992) 2019.

\bibitem{stokstad}
R.G. Stokstad, Y. Eisen, S. Kaplanis, D. Pelte, U. Smilansky and I. Tserruya,
Phys. Rev. C 21 (1980) 2427.

\bibitem{jack1}
J.X. Wei {\it et al.}, Phys. Rev. Lett. 67 (1991) 3368.

\bibitem{jack2}
J.R. Leigh {\it et al.}, Phys. Rev. C 47 (1993) R437.

\bibitem{jack3}
J.R. Leigh {\it et al.}, in: Proc. RIKEN workshop on Heavy-Ion
Reactions with Neutron-Rich Beams (Japan, 1993).

\bibitem{wuosmaa}
A. H. Wuosmaa {\it et al.}, Phys. Lett. B 263 (1991) 23.

\bibitem{bierman}
J.D. Bierman, A.W. Charlop, D.J. Prindle, R. Vandenbosch and D. Ye, Phys. Rev.
C 48 (1993) 319.

\bibitem{baba}
C. V. K. Baba, Nucl. Phys. A 553 (1993) 719c.

\bibitem{nds}
Nuclear Data Sheets 58 (1989) 93; Nuclear Data Sheets 59 (1990) 393.

\bibitem{argentina}
D.E. DiGregorio {\it et al.}, Phys. Lett. B 176 (1986) 322.

\bibitem{stokgross}
R.G. Stokstad and E.E. Gross, Phys. Rev. C 23 (1981) 281.

\end{thebibliography}
\end{document}